# Financing Urban Infrastructure Through Land Leasing: Evidence From Bahir Dar City, Ethiopia

Wudu Muluneh[1] and Tadesse Amsalu[1] ,

[1] Institute of Land Administration, Bahir Dar University, Bahir Dar, Ethiopia

**Abstract**

The provision of essential urban infrastructure and services for the expanding population is a persistent financial challenge for many of the rapidly expanding cities in developing nations like Ethiopia. The land lease system has received little academic attention as a means of financing urban infrastructure in developing countries. Therefore, the main objective of this study is to assess the contribution of land leasing in financing urban infrastructure and services using evidence from Bahir Dar city, Ethiopia. Primary and secondary data- gathering techniques have been used. Descriptive statistics and qualitative analysis have been adopted. The results show land lease revenue is a dominant source of extra-budgetary revenue for Bahir Dar city. As evidenced by Bahir Dar city, a significant portion of urban infrastructure expenditure is financed by revenues from land leasing. However, despite the critical importance of land lease revenue to investments in urban infrastructure, there is inefficiency in the collection of potential lease revenue due to weak information exchange, inadequate land provision for various uses, lack of transparency in tender committees, and the existence of poor documentation. Our findings suggest that Bahir Dar City needs to manage lease revenue more effectively to increase investment in urban infrastructure while giving due consideration to availing more land for leasing.

1. **Introduction**

Urbanization is defined in different ways from different perspectives. Urbanization is defined as a global phenomenon [1], which refers to a growth in the proportion of a population living in urban areas and the further physical expansion of urban centers that already exist [2]. It is also referred to as a sophisticated socio-economic process that alters the built environment, turns rural communities into urban settings, and also shifts the spatial distribution of a population from rural to urban areas[1] [3].

The comparison of the present level of urbanization figures with historical trends helps to determine the pace of urbanization. About 55.2 % of the world's population is confined to cities as of 2018 [3]. According to projections, 68% of people will reside in urban areas by 2050 [3]. The increase will primarily come from Asia and Africa at about 90% [3]. Africa is currently experiencing a 43% urbanization level [3]. Sub-Saharan Africa experienced an annual urban population growth rate of 4.1 percent between 2000 and 2015 [4]. The annual population growth rate in East Africa is even higher, at 6.5 % [5]. When compared to a global rate of 2.0%, this is extremely high. About 40 % of Africa's urban population growth is a result of rural-to-urban migration [6]. Between now and 2050, the rate of urbanization in Africa is anticipated to increase significantly [5] as the continent tries to catch up with the rest of the world.

---
[1] This is the definition we carry forward

Ethiopia has a 21.2% urbanization rate as of 2019[3]. Even when compared to SSA countries, this is very low [7]. However, Ethiopia is a country experiencing fast urbanization. The urban population increased by 4.2 % per year between 1994 and 2015[7]. The annual growth rate has surpassed 6.2 % since 2011 [8]. According to projections, the nation's urban population will make up 39% of the total population by 2050 [3]. However, small towns and intermediate cities are where the majority of Ethiopia's urbanization is occurring [9, 10]. This pattern will continue for the next 20 years [7].

In the context of rapid world urbanization, the demand for urban infrastructure and services is growing from time to time [11]. To ensure socioeconomic vitality and reduce social inequality during the urbanization processes, adequate investment in urban infrastructure systems is therefore required [12]. But during periods of rapid urbanization, cities throughout the developing world have faced significant difficulties in providing enough capital investment in urban infrastructure [12, 13]. For instance, between 23 and 65 % of the urban population in developing nations resides in slums that lack basic water supply infrastructure, sanitary facilities, and permanent shelters [14].

In Ethiopia, there is unprecedented increase in demand for urban infrastructure and services, including water and sanitation, safe disposal of wastes, affordable housing, well-connected transport systems, and other infrastructures and basic services, both in major and smaller cities [15]. Inadequate maintenance, unmet demand, and problems with cost recovery make it difficult for the public sector to provide it more effectively [15]. As a result, there is a serious lack of infrastructure in many Ethiopian municipalities. For instance, in Bahir Dar city, 31% of its residents are in absolute poverty that do not have access to safe water supply, sanitation (public and household toilet facilities), and drainage facilities[16]. In addition, inadequacy of the road infrastructures, green areas and abattoir facilities are fundamental problems of the city [16].

Numerous previous researches indicated that one of the bottlenecks that African countries face including Ethiopia is lack of financial resources to address infrastructure backlog [11, 15, 17]. Traditional revenue sources, which African local governments depend on, tend to be inefficient and minimal as a result they remain unable to shape their own development and deliver much-needed infrastructure services [18, 19]. Above everything else scholars indicated that there is no one best way of funding mechanism for infrastructure; hence depending on their institutional, political, economic, cultural and social context, countries should look for the most effective alternative financing options of infrastructure [20].

In a nation where the land resource is owned by the government, land is the most valuable resource for funding municipal services [21]. For instance, land leasing is regarded as a tool for financing infrastructure investments in Chinese cities [21, 22]. Between 45 and 80 % of all local government revenues in Guangdong province and Shenzhen cities in the 1990s and roughly 60 to 80 % of local capital expenditures in Beijing and Chengdu from 1995 to 2002 came from land leasing [21].

A great deal of studies in Ethiopia examined the land lease system as prodigious way of legitimizing the transfer of land use rights from the government to individual citizens and its role as a means to meet the increasing demand for more urban land. However, the available literature has given little attention to examining its contribution in generating revenue for urban infrastructure provision [23-27]. Hence, we know very little about the role of the land lease system in generating revenue over the years to provide basic infrastructure. Consequently, it is deemed essential to ascertain the role of land leasing in providing the financial support needed to meet the increased demand for infrastructure.

Therefore, the purpose of this study is to examine the contribution of land leasing revenue for financing urban infrastructure from empirical evidences gathered at Bahir Dar City, Ethiopia.

The paper is organized into 5 sections. Section 1 presents the purpose of the paper while sections 2 focuses on the literature review and in sections 3 methods of the study are addressed. Results and discussion are addressed under section four. Finally, the conclusion is presented in section 5 to summarize the findings and highlight policy implications.

## 2. Literature Reviews

This section reviews the basics of capturing land values, leasing land, and urban infrastructure or services. It presents and analyzes the scholarly debate surrounding such concepts, concentrating on their linkages and important characteristics in the context of this study.

### 2.1 The concept of Land Leasing

The term leasing was mostly used to refer to real estate rentals until the 1950s, when it was expanded to include movable properties [28]. Due to the confusion created by this new dimension in defining leases for movable and immovable property, the term land lease was used to refer solely to the lease of real property or land [28]. Because the land leasing system is directly dependent on the land tenure systems, it is critical to understand land leasing from that standpoint. Among the rights to land or property are freehold (fee simple), leasehold, freehold ground rent, and mortgage [29]. A freehold interest in land entails full financial risk of ownership and can be held by an owner occupier or an investor. In other terms, it is a perpetual unrestricted interest. Leasehold refers to the transfer of some of the freeholder's rights for a set amount of time (fixed years) in exchange for a capital payment (premium) and/or a fixed monthly income (ground rent). The annual payment of long leases for a set amount of time under a freehold ground rent arrangement, on the other hand, is typically made on undeveloped land. However, it can be leased at developed sites for less than the value of the underlying real property (land and improvements).A mortgage is a long-term monetary borrowing secured by property[29].

The global land tenure system includes customary tenure, private tenure, public tenure, religious land tenure, and non-formal tenure. Except for non-formal tenure, which has no legal support, leasehold can be produced from any of the aforementioned forms of tenure. As a result, the terms "customary," "public," "private," and "religious" are used to refer to the respective types of leases derived from customary, public, private, and religious tenures. Therefore, the word land leasing refers to the embodiment of any of these agreements where a lease is provided from any of these groups. No matter what kind of lease category it falls under, there are some common features that apply, including a set number of years, the option to renew after it expires, the transfer of land rights, and payment methods, such as a one-time or annual payment [30,31]. A lease's duration is flexible and differs from country to country.

Based on payment mode, the public leasehold system can be divided into two parts: premium (also known as leasing fee) and land rent [30]. Contrary to the term premium system, which refers to when a lump sum or the capital value of the land is paid for the duration of the lease, the land rent system refers to when the lessee pays the lease on an annual basis only. Canberra, Australia, Israel, and China, for example, employ the premium system, whereas Ukraine, Russia, and Sweden use the land rent system [30, 31]. However, there are circumstances where both exist [29,32]. Examples of countries with premium and land rent systems include Ghana and the Netherlands.

All land in Ethiopia is owned by the government through the public leasehold system that allows for the acquisition of land. The lease system was initially put into place in 1993 (Proc. 80/1993), was later repealed (Proc. 272/2002), and was finally replaced (Proc. 721/2011). For a predetermined length of time (99 years for residential, 90 years for institutions, 70 years for businesses, and 60 years for other uses), land in urban areas is leased to individuals and developers. Ethiopia does not need the full amount of the land lease up front. A portion of lease payment is due at contract signing. The remainder can be paid over the lifetime of the lease agreement.

## 2.2. Land Leasing as a Land Value Capture Instrument

Land leasing is a tool for capturing land values, particularly in public land leasehold systems [30]. Premiums or ground rents can be used to recover land values attributed to the community as a whole [30,31]. The potential is not fully realized because the premium or ground rent is relatively low in comparison to the value of the land [30,31]. This is largely because, in practice, land rent or ground rent is sometimes misinterpreted as land tax, particularly in the public leasehold system [31]. Land tax, unlike ground rent, is a type of property tax (land value tax) used to fund public infrastructure. This difficulty does not exist in the other types of leasehold because the landowner (freeholder) is separate from the taxation authority, as opposed to public land leasing, where the state is both the landowner and the taxing authority. Land leasing is a good value capture mechanism in China and it helps in financing urban infrastructure, promoting economic development, and promoting equal distribution and access to land [30, 33].

## 2.3 Urban Infrastructure and Services and Economic Growth

The relationship between infrastructure development and economic growth has been the subject of numerous studies recently. Empirical studies in Pakistan reveal a strong correlation between development of the physical environment and economic expansion. The study's conclusions showed how important infrastructure building is to fostering and sustaining economic growth over the long term. The major factors that determine economic growth, namely physical infrastructure, can also be improved by investing in the economy [34].

A study on the effects of infrastructural development on residential property values in Minna-Nigeria indicates the provision of adequate infrastructure is a tool for driving economic development within a given locality [35]. A study in Indonesia found an asymmetric relationship between infrastructure and economic growth in the long run, but a symmetric relationship in the short run. Furthermore, this study uncovers the causal direction of Indonesian economic development, which shifts from gross fixed capital formation to labor [36]. The paper recommends increasing investment in the infrastructure sector to help the Indonesian economy grow. The study also encourages policymakers to develop robust infrastructure policies that guide infrastructure and the country's economy in both the short and long run [36].

Empirical evidence from China suggests that new infrastructure investment can significantly improve economic growth quality, and this conclusion appears to be valid after a serious of researches. Studies show that new infrastructure investments help to improve economic growth quality in terms of the condition, process, and outcomes of economic growth by encouraging technological innovation, improving industrial structure, and increasing production efficiency [37].

Studies on the effects of infrastructure development on growth and income distribution over 100 countries and spanning the years 1960-2000 shows growth is positively affected by the stock of infrastructure assets and infrastructure development can be highly effective to combat poverty [38].

## 2.4. Urban Infrastructure and Land Values

The term infrastructure refers to the physical structures that support a society, such as roads, water supply, sewers, electrical national grids, telecommunications, and so forth, and can be defined as "the physical components of interrelated systems providing commodities and services essential to enable, sustain, or enhance societal living conditions" [39].

Public investments, particularly in infrastructure and services, have primarily influenced land values and many scholars have argued in favor of the positive impact on land values, while a few have argued in favor of neutral or negative consequences [40]. For example, empirical evidence on the effects of infrastructure and property development on surrounding land values in selected neighborhoods in Greater Port Harcourt City of Rivers State revealed that the presence of infrastructure facilities such as tarred roads, drainage and services, results in increases in land values. Active property development also affected land values and brought about an upward trend in value of land and landed properties within the studied neighborhoods [39]. The findings also reveal that lack of infrastructure retards property development and led to stagnation or even decline in land values [39].Other studies also showed that areas with basic facilities such as access roads, good drainage, electricity, public water supply and telephone attract high property values [41]. Another study also presented accessibility which is a direct consequence of a good road network, in turn leads to high rental values of locations with greatest accessibility advantages [42]. In a situation where all properties are accessible via motorable road network, such properties will enjoy high rental values conferred by virtue of accessibility.

## 3. Materials and Methods

### 3.1 Study Area Description

This study was conducted in Bahir Dar city, Amhara Region, Ethiopia. Bahir Dar is the capital city of Amhara National Regional State. According to the Amhara National Regional State proclamation 91/2003, Bahir Dar has a metropolitan city grade status since 2003. The population of Bahir Dar is estimated to be 440,847 in 2021 [16]. The current master plan of Bahir Dar city encompasses four satellite kebeles (the lowest level administrative set up) namely Meshenti, Zenzelima, Zege and Tis Abay located immediately after the municipal boundaries.

Bahir Dar was specifically chosen for the case study because it was the first in the region to implement a land lease system [43]. The other reason for choosing Bahir Dar city is that despite the city's efforts to meet rising demand for infrastructure and services, there is still a significant gap in provision, as indicated in the introduction section. As a result, understanding how the city uses lease revenue to fund infrastructure will serve as a model for other cities.

### 3.2 Research method, data collection and analysis

To meet the research objective, the approach applied in this study is a mixed approach, which enabled to minimize the shortcomings of using a specific approach and the shortcomings of accessing quantitative data.

Relevant quantitative and qualitative data for the study was collected both from primary and secondary sources. As to the firsthand information, questionnaire was used to collect data from a total of 25 purposely selected experts from the Bahir Dar city administration, which included 15 employees from the land development and management office and 10 employees from the Bureau of revenue collection. In addition seven key informants (three from the land administration department of the Municipality and four from Bahir Dar city revenue offices) were selected on purpose based on their academic status and years of experience to explain the issues and challenges in land leasing revenue mobilization and collection. Experts with high academic status and years of experience were given priority. Furthermore, in-depth interview / focus group discussion was held with 6 land leasing experts from the lease department of the city administration to clarify the information or data collected during key informant interviews.

Secondary data have been collected analyzed by reviewing different published and unpublished documents such as urban land development and management policy document, lease proclamation, lease regulation, and directives, annual plan and reports of the city administration among others. The collected data have been summarized and presented in tables, figures and charts using appropriate application (SPPS and MS excel). Descriptive statistics such as percentages, mean, minimum and maximum and the likes are used to analyze the data.

## 4. Results and Discussion

### 4.1 Land Leasing Revenue Collection Performance

The city of Bahir Dar has land leasing authority. Land leasing is one of the main sources of revenue for the city. The revenue collected from land lease includes administration costs, which consists of yearly land rent, an upfront lease payment, and the sale of tender documents. The city gets significant inflows from leases of urban lands, which allows for the wide range of development of infrastructure and service demands. Table 1 indicates revenue collected from leasing during 2017 to 2021. The minimum annual land lease revenue collection efficiency was recorded in 2021; whereas the plan was to collect 316,000,000 Birr, only 45.9% (144,905,949.28 Birr) was collected. The maximum collecting efficiency was recorded in 2017; the plan was 160,000,000 Birr and the actual collected was 194,269,842 Birr (i.e. a 127% achievement). Although, the average collection performance during the stated periods is 87%, the growth of annual revenue collection indicates irregular trend.

While the accomplishment is remarkable, key informants were asked about why the city was unable to fully implement its plan. They stated that "the collection efficiency of the land lease was poor due to weak information exchange, inadequate provision of land for different uses, problems in tender committees, as well as the existence of poor documentation." On the other hand, similar studies in China showed that due to rapid urbanization and the boom in the real estate sector, land lease revenue has shown dramatic increases since the early 2000s[21]; and the revenue has played a crucial role in financing local governments in the form of lump sum revenue and/or collateral to undertake debt financing [46]. Land lease revenue was 5.95 billion Yuan in 2000, and It increased steadily thereafter, reaching 437.45 billion Yuan, in 2013[21].

Table 1: - Land lease revenue collection performance of Bahir Dar City 2017-2020

| Year | Plan (Birr) | Actual (Birr)[2] | Actual Vs Plan | Growth Rate |
| --- | --- | --- | --- | --- |
| 2017 | 160,000,000 | 194,269,842 | 121.4% | |
| 2018 | 220,000,000 | 204,747,900 | 93.1% | 6% |
| 2019 | 190,000,000 | 175,452,757 | 92.3 | -14% |
| 2020 | 250,000,000 | 271,750,370.08 | 109% | 55% |
| 2021 | 316,000,000 | 144,905,949.28 | 45.9% | -46% |
| Average | 227,200,000 | 198,225,363.67 | 87.2% | |

Source: Data collected from Bahir Dar City Administration Revenue Enhancement Plan

### 4.2 The Share of Land Lease Revenue in The Overall Municipal Revenue

Municipal revenues are those that the city administration collects from local citizens and local business, and assumed to be amalgamated with regional transfers. These include tax from business and professional service tax, rental income tax, annual vehicle tax, urban land lease, urban land rent, recreational centers, rent from residual house, engineering service, issuance of fence buildings permit, issuance of certificate etc [44].

Table 2 depicts Bahir Dar city Administration's land lease revenue as a percentage of municipal revenue collected between 2010/11 and 2020/2021. As depicted in Table. 2, land lease revenue increased steadily from 21,382,377.77 in 2010 to 245,223,268.5 in 2020/21. The percentage of land lease revenue in municipal revenue fluctuated from 40.3% to 71.6% during this period with average of 56%. The remaining 46% is constituted by other sources of revenue. Thus, the lion share of the municipal revenue is constituted by land lease. This is in conformity with a research finding that states leasing has become the single largest source of municipal revenue in Ethiopia, overtaking the traditionally largest source of revenue, the local fee and tax items covered in Regional Tariff Proclamations' [20]. Specifically, the study pointed out that the revenue from land lease ranges from 21% to 45% of total revenue of the cities in Ethiopia [20]. Similarly approximately 45% and 80% of total local government revenues in Guangdong province and Shenzhen cities of China, in the 1990s were contributed from land leasing respectively [21]. The authors assert that the heavy dependence of the city on land lease revenue may hurt sustainable urban development as land is a scarce resource.

---
[2] Birr is the medium of exchange in Ethiopia. One US dollar equals 60 Birr.

Table 2: Bahir Dar City Administration Land Lease Revenue Proportion Out of the total Municipal Revenue, 2010/11 - -2020/21

| Year | Land Lease Revenue (Birr) | Total revenue (Birr) | Land lease revenue proportion out of total revenue in % |
|---|---|---|---|
| 2010/11 | 21,382,377.77 | 40,879,000 | 52.31 |
| 2011/12 | 27,082,360.98 | 67,204,246.36 | 40.30 |
| 2012/13 | 84,262,557 | 117,754,036.49 | 71.60 |
| 2013/14 | 99,483,265.39 | 149,655,637 | 66.50 |
| 2014/15 | 108,570,302.90 | 179,934,144.94 | 60.34 |
| 2015/16 | 110,311,694.15 | 199,249,377.26 | 55.40 |
| 2016/17 | 194,269,842.07 | 326,442,216.17 | 59.51 |
| 2017/18 | 204,747,900.63 | 361,726,357.61 | 56.60 |
| 2018/19 | 175,452,757.01 | 368,648,216.01 | 47.60 |
| 2019/20 | 271,750,370.08 | 438,994,047 | 61.9 |
| 2021/22 | 144,905,949.28 | 414,952,958.88 | 34.9 |
| Average | | | 55.17 |

Sources: Bahir Dar City Administration three year Capital Investment Plan (CIP) and Revenue Enhancement Plan (REP) Annual Report (from 2011/12 – 2018/19)

### 4.3 Proportion of Land Leasing Revenue Spent on Urban Infrastructure Investment

The experience on the proportion of land leasing revenue spent on urban infrastructure varies from country to country. For instance, in Hong Hong, and India, most of the revenue traditionally has been used to pay for infrastructure investment and other public works, although there is no legal earmarking of land-sale proceeds [20,45]. Similarly, experiences in Singapore, Australia, London, and Stockholm show that the lease system is a major source of revenue for financing infrastructure and other urban development projects such as housing, transportation, and utilities. However, in Ethiopia the land lease policy attempts to tie land-leasing revenue directly to municipal infrastructure investment. The federal land-leasing proclamation states that a municipality shall earmark 90 percent of all land-leasing proceeds for infrastructure investment [21]. The remaining 10% is allocated to administrative tasks. Bahir Dar city serves as an example of this. Key informants claim that 90% of municipal land leasing revenues are always used for capital projects, with the remaining 10% going toward administrative costs.

### 4.4 The Share of Land Lease Revenue From Total Capital Expenditure

Investment in urban infrastructure plays a significant role in advancing urbanization and urban development. In Bahir Dar, the capital expenditure increased steadily from 60, 200,678 Birr in 2010/11 to 547,121, 925 Birr in 2020/21(Figure 1). The percentage of capital expenditure covered by land lease revenue fluctuated from 30% to 68% during this period with an average of 44 %. The capital expenditure of the municipality is normally concerned with the creation of long term assets such as construction of roads, drainage, slaughter houses, housing and other social infrastructure development of the city.

Figure 1: Bahir Dar City Administration Land Lease Revenue and capital expenditure, 2010/11 - 2020/21

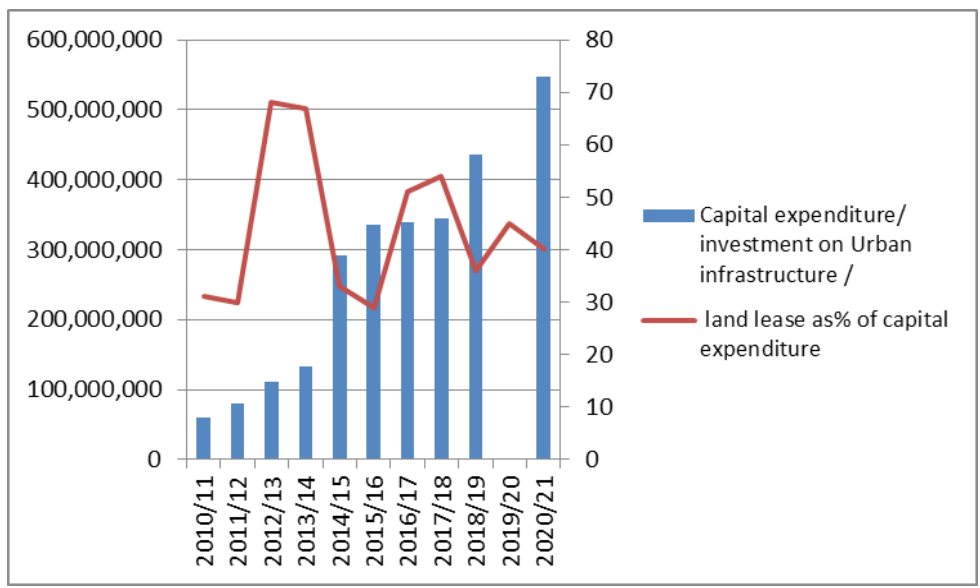

Source: Bahir Dar City Administration

The fact on the ground leads to daringly argue that land leasing revenue provides financial support for urban infrastructure investments to promote urbanization through the development of socio-economic, cultural, political, and environmental factors of the city. The above findings are in line with the views of scholars who contend that public land revenues should be used for investments in urban infrastructure [33]. Evidences showed that many cities in China have financed half or more of their very high urban infrastructure investment levels directly from land leasing, and the remaining covered by borrowing from financial institutions using their land as collateral [23].

### 4.5 Challenges in Municipal Lease Revenue Mobilization

Bahir Dar city face a rapidly growing demand for urban infrastructure and services as a result of continuing rapid population growth. Its capacity to supply urban infrastructure and services is severely constrained by a shortage of fiscal resources through a sound revenue generation base. Bahir Dar city not only encountered with a massive backlog of new infrastructure requirements but also needs to allocate more resources for maintenance, renovation and replacement of older and dilapidated equipment. With this background, the challenges the city faces in land lease revenue mobilization are the following.

#### 4.5.1. Institutional Inefficiency in Land Lease Revenue Collection

The focus group discussion implied that one of the reasons for inefficiency in land lease revenue mobilization is the lack of coordination among revenue office and other responsible sectors in identification and registration of lease payers. There is a gap in data and information exchange between the revenue office and the land administration office. The other problem pointed out is inadequate provision of land for different uses and problems in tender committees as well as the existence of poor documentation. Furthermore, there is weak follow up and monitoring mechanism to collect the lease revenue as per the lease agreement. Hence, a great deal of the lease revenue remains in arrears that should have been collected. The potential amount of money that could be collected from land lease is being planned incorrectly, which is the other technical cause of inefficient capacity utilization. Since the plans are lower than the city administration's potential lease revenue, it would have been possible to stretch the budget so that the city administration could use all of its available resources effectively and efficiently.

### 4.5.2. Inefficient Land Value Capture

The potential lease income from land has not yet been fully realized in Bahir Dar because demand and supply forces do not operate there to determine land value. There is no obvious way for the city to take advantage of the increase in land value in areas that have become more desirable and livable as a result of investments in municipal infrastructure and services. The value of a piece of land in Bahir Dar is calculated by multiplying its size by the lease rate specified in the lease directive. The transfer of land, however, does not take into account the current market value because the rate is not periodically adjusted to reflect the state of the market.

### 5. Conclusion

The primary goal of this study was to assess urban land lease revenue as an option for financing urban infrastructure investment by taking evidence from Bahir Dar city. The study found that land lease revenue has become a significant source of revenue for Bahir Dar city municipality that allows the opportunity to increase capital investments and, as a result, boost economic growth. As evidenced by the city of Bahir Dar, a considerable portion of urban infrastructure investment expenditure is covered from land lease revenue.

Although land lease revenue is critically significant for financing urban infrastructure investment, there is inefficiency in the collection of potential lease revenue due to weak information exchange, inadequate land provision for various uses, lack of transparency in tender committees, and the existence of poor documentation. Our findings suggest that Bahir Dar City needs to manage lease revenue more effectively to increase investment in urban infrastructure.

Funding: This research did not receive any specific grant from funding agencies in the public, commercial, or not-for-profit sectors